\documentclass[preprint,showpacs,preprintnumbers,amsmath,amssymb]{revtex4}

\usepackage{graphicx}% Include figure files
\usepackage{dcolumn}% Align table columns on decimal point
\usepackage{bm}% bold math
\usepackage{multirow}
\begin{document}

\title{General Relativistic Aberration Equation and Measurable
Angle of Light Ray in Kerr--de Sitter Spacetime}% Force line breaks with \\

\author{Hideyoshi ARAKIDA}
\email[E-mail:]{arakida.hideyoshi@nihon-u.ac.jp}
\affiliation{College of Engineering, Nihon University,
Koriyama, Fukushima 963-8642 JAPAN}

\date{\today}% It is always \today, today,
             %  but any date may be explicitly
\begin{abstract}
As an extension of our previous paper, instead of the total deflection
angle $\alpha$, we will mainly focus on discussing the measurable
angle of the light ray $\psi$ at the position of the observer in
Kerr--de Sitter
spacetime which includes the cosmological constant $\Lambda$.
We will investigate the contributions of the radial and transverse
motions of the observer which are connected with the radial velocity
$v^r$ and transverse velocity $bv^{\phi}$ ($b$ is the impact parameter)
as well as the spin parameter $a$ of the central object
which induces the gravitomagnetic field or frame dragging
and the cosmological constant $\Lambda$.
The general relativistic aberration equation is employed
to take into account the influence of the motion of the observer
on the measurable angle $\psi$. The measurable angle $\psi$ derived
in this paper can be applied to the observer placed within
the curved and finite-distance region in the spacetime.
The equation of the light trajectory will be obtained in such a way that
the background is de Sitter spacetime instead of Minkowski spacetime.
If we assume that the lens object is the typical galaxy,
the static terms ${\cal O}(\Lambda bm, \Lambda ba)$ are basically
comparable with the second order deflection
term ${\cal O}(m^2)$, and they are almost one order smaller that
of the Kerr deflection $-4ma/b^2$. The velocity-dependent terms
${\cal O}(\Lambda bm v^r, \Lambda bav^r)$ for radial motion and
${\cal O}(\Lambda b^2mv^{\phi}, \Lambda b^2av^{\phi})$ for
transverse motion are at most two orders of magnitude smaller
than the second order deflection ${\cal O}(m^2)$.
We also find that even when the radial and transverse velocities have
the same sign, their asymptotic behavior as $\phi$ approaches $0$ is 
differs, and each diverges to the opposite infinity.
\end{abstract}


\maketitle

%\tableofcontents

%%%%%%%%%%%%%%%%%%%%%%%%%%%%%%%%%%%%%%%%%%%%%%%%%%%%%%%%%%%%%%%%%%%%
\section{Introduction\label{sec:intro}}

The cosmological constant problem is an old but unsolved issue
in astrophysics and cosmology that is closely related to
the general theory of relativity; see reviews by, e.g.,
\cite{weinberg1989,carroll2001}.
After the establishment of the general theory of relativity
in 1915--1916, Einstein incorporated the cosmological term
$\Lambda g_{\mu\nu}$ into the field equation in order to represent
the static Universe. Although the discovery of cosmic expansion by
Hubble caused Einstein to withdraw
the cosmological term from the field equation, nowadays it is
widely considered that the cosmological constant $\Lambda$,
or the dark energy in a more general sense, is the most promising
candidate for explaining the observed accelerating expansion of
the Universe \cite{riess1998,schmidt1998,perlmutter1999}
despite the fact that its details are not at all clear.
One straightforward way to tackle this problem from another viewpoint
is to investigate the effect of
the cosmological constant $\Lambda$ on the bending of a light ray.
In fact, the bending of a light ray is the basis of gravitational
lensing which is a powerful tool used in astrophysics and
cosmology; see, e.g., \cite{schneider_etal1999,schneider_etal2006}
and the references therein.

The influence of the cosmological constant $\Lambda$ on
light deflection, especially on the total deflection angle $\alpha$,
had been the subject of a long debate and was investigated mainly
under the static and spherically symmetric vacuum solution,
namely the Schwarzschild--de Sitter/Kottler solution.
Islam \cite{islam1983} first showed
that the trajectory of a light ray is not related to
the cosmological constant $\Lambda$ because the second-order
differential equation of the light ray does not depend on $\Lambda$.
On the basis of the result obtained by Islam, it was thought for
a long time that the cosmological constant $\Lambda$ does not affect
the bending of a light ray. However, in 2007, a significant indication was
provided by Rindler and Ishak \cite{rindler2007} who pointed out
that the cosmological constant $\Lambda$ {\it does} contribute to
the bending angle of a light ray in terms of the invariant cosine
formula. The important point of \cite{rindler2007} is that though
the trajectory equation of the light ray admittedly does not depend on
the cosmological constant $\Lambda$, the angle should be determined
via the metric tensor $g_{\mu\nu}$: because Schwarzschild--de
Sitter spacetime is not asymptotically flat, the metric $g_{\mu\nu}$
plays an important role in determining the angle as well as the length
\footnote{
If one obtains the deflection angle using the equation of
the light trajectory {\it only}, it means that the angle is evaluated
in flat spacetime because only in flat spacetime,
do $r$ and $\phi$ have the meaning of the length and angle, respectively.
On the other hand, in curved spacetime, $r$ and $\phi$ are just
``the coordinate values'' and then the angle and length should be determined
by the metric.}.
Inspired by this paper, many authors intensively discussed its
appearance in diverse ways;
see \cite{ishak2010} for a review article, and also see, e.g.,
\cite{lake2002,park2008,kp2008,sph2010,bhadra2010,miraghaei2010,biressa2011,ak2012,lebedevlake2013,hammad2013,batic_etal2015,arakida2016,ishihara_etal2016,arakida2018a}
and the references therein. Moreover, several authors discussed
light deflection in Kerr--de Sitter spacetime which is
the stationary and axially symmetric vacuum solution and
includes the spin parameter $a$ of the central object
as well as the cosmological constant $\Lambda$; see, e.g.,
\cite{kraniotis2005,kraniotis2011,sultana2013,charbulak_stuchlik2017}
and the references therein. The same consideration is further extended
to the more general Kerr-type solutions, see, e.g.,
\cite{goicoechea_etal1992,iyer_hansen2009,kraniotis2014,he_lin2016,he_lin2017a,he_lin2017b,jiang_lin2018,uniyal_etal2018}.

However, in spite of the intensive discussions and various approaches,
a definitive conclusion has not yet emerged. One of the main
reasons for this is that because the Schwarzschild--de Sitter and
Kerr--de Sitter solutions are not asymptotically flat unlike
the Schwarzschild and Kerr solutions, it is ambiguous and unclear
how the total deflection angle $\alpha$ should be defined in
curved spacetime. Although to overcome this difficulty a method
for calculating the total deflection angle is independently
investigated and proposed on the basis of the Gauss--Bonnet theorem by
\cite{ishihara_etal2016,arakida2018a},
it seems that further consideration is needed to settle the argument.
Because of the difficulty of defining the total deflection angle
$\alpha$ in curved spacetime, our argument is described in
Appendix A of \cite{arakida2018b}.

As described briefly above, the concept and definition of the total
deflection angle of the light ray $\alpha$ is a counterintuitive
and difficult problem; however it is always possible to determine
the measurable angle $\psi$ at the position of the observer $P$
which can be described as the intersection angle between the tangent
vector $k^{\mu}$ of the light ray $\Gamma_k$ that we investigate
and the tangent vector $w^{\mu}$ of the radial null geodesic $\Gamma_w$
connecting the center $O$ and the position of observer $P$.
See FIG. \ref{fig:arakida-fig1} in section \ref{sec:aberration}.
We investigated in \cite{arakida2018b} the measurable angle of the
light ray $\psi$ at the position of the observer in the Kerr spacetime
on the basis of the general relativistic aberration equation
\cite{lebedevlake2013}
(and see also \cite{pechenick_etal1983,lebedevlake2016})
which enables us to compute the effect of the motion of the observer
more easily and straightforwardly because the equations of the null
geodesic of $\Gamma_k$ and $\Gamma_w$ do not depend on the motion of
the observer; the velocity effect is incorporated in the formula
as the form of the 4-velocity of the observer $u^{\mu}$.

In present paper, we will extend our discussion of the measurable
angle $\psi$ at the position of observer $P$ to Kerr--de Sitter
spacetime containing the cosmological constant $\Lambda$ as well as
the spin parameter of the central object $a$. Our purpose is to
examine not only the contribution of the cosmological constant
$\Lambda$ and the spin parameter $a$ of the central object but also
the effect of the motion of the observer on the measurable angle $\psi$.
As in our previous paper, the 4-velocity of the observer
$u^{\mu}$ is converted to the coordinate radial velocity $v^r = dr/dt$
and coordinate transverse velocity $bv^{\phi} = bd\phi/dt$
($b$ is the impact parameter and $v^{\phi} = d\phi/dt$ denotes
the coordinate angular velocity), respectively.

This paper is organized as follows: in section \ref{sec:trajectory},
the trajectory of a light ray in Kerr--de Sitter spacetime is derived
from the first-order differential equation of the null geodesic.
In section \ref{sec:aberration}, the general relativistic
aberration equation is introduced and in section
\ref{sec:angle} the measurable angle $\psi$ in Kerr--de Sitter
spacetime is calculated for the cases of the static observer,
the observer in radial motion and the observer in transverse motion.
Finally, section \ref{sec:conclusions} is devoted to presenting
the conclusions.

%%%%%%%%%%%%%%%%%%%%%%%%%%%%%%%%%%%%%%%%%%%%%%%%%%%%%%%%%%%%%%%%%%%%
\section{Light Trajectory in Kerr--de Sitter Spacetime\label{sec:trajectory}}
The Kerr--de Sitter solution --- see Eqs. of (5.65) and (5.66) in
\cite{carter1973}, and also, e.g.,
\cite{kraniotis2005,kraniotis2011,sultana2013,charbulak_stuchlik2017} ---
in Boyer--Lindquist type coordinates $(t, r, \theta, \phi)$ can be
rearranged as
\begin{eqnarray}
 ds^2 &=& g_{\mu\nu}dx^{\mu}dx^{\nu}
  \nonumber\\
 &=&
  - \frac{\Delta_r - \Delta_{\theta} a^2 \sin^2\theta}{\rho^2\Xi^2}dt^2
  + \frac{\rho^2}{\Delta_r}dr^2
  + \frac{\rho^2}{\Delta_{\theta}}d\theta^2
  \nonumber\\
 & &- \frac{2a\sin^2\theta}{\rho^2\Xi^2}
  \left[
   \Delta_{\theta}(r^2 + a^2) - \Delta_r
		\right]dtd\phi
  \nonumber\\
 & &+ \frac{\sin^2\theta}{\rho^2\Xi^2}
  \left[
   \Delta_{\theta}(r^2 + a^2)^2 - \Delta_r a^2 \sin^2\theta
  \right]d\phi^2,
  \label{eq:kerr-deSittermetric1}
\end{eqnarray}
where
\begin{eqnarray}
 \Delta_r
  &=& r^2 + a^2 - 2mr - \frac{\Lambda}{3}r^2(r^2 + a^2),
  \label{eq:delta_r}\\
 \Delta_{\theta}
  &=&
  1 + \frac{\Lambda}{3}a^2\cos^2\theta,
  \label{eq:delta_theta}\\
 \rho^2
  &=& r^2 + a^2\cos^2\theta,
  \label{eq:rho}\\
 \Xi
  &=& 1 + \frac{\Lambda}{3}a^2.
  \label{eq:Xi}
\end{eqnarray}
$g_{\mu\nu}$ is the metric tensor; Greek indices, e.g., $\mu, \nu$,
run from 0 to 3; $\Lambda$ is the cosmological constant; $m$ is
the mass of the central object; $a \equiv J/m$ is a spin parameter
of the central object ($J$ is the angular momentum of the central
object) and we use the geometrical unit $c = G = 1$
throughout this paper.

For the sake of brevity, we restrict the trajectory of the light ray
to the equatorial plane $\theta = \pi/2, d\theta = 0$, and rewrite
Eq (\ref{eq:kerr-deSittermetric1}) in symbolic form as
\begin{eqnarray}
 ds^2 = -A(r)dt^2 + B(r)dr^2 + 2C(r)dtd\phi + D(r)d\phi^2,
  \label{eq:kerr-deSittermetric2}
\end{eqnarray}
where $A(r), B(r), C(r)$, and $D(r)$ are
\begin{eqnarray}
 A(r) &=& \left(1 + \frac{\Lambda}{3}a^2\right)^{-2}
  \left[
   1 - \frac{2m}{r} - \frac{\Lambda}{3}(r^2 + a^2)
  \right],
  \label{eq:Ar}\\
 B(r) &=&
  \left[
   \left(1 - \frac{\Lambda}{3}r^2\right)
   \left(1 + \frac{a^2}{r^2}\right) - \frac{2m}{r}
  \right]^{-1},
  \label{eq:Br}\\
 C(r) &=&
  - \left(1 + \frac{\Lambda}{3}a^2\right)^{-2}
  a\left[\frac{2m}{r} + \frac{\Lambda}{3}(r^2 + a^2)\right],
  \label{eq:Cr}\\
 D(r) &=&
  \left(1 + \frac{\Lambda}{3}a^2\right)^{-2}
  \left[
   (r^2 + a^2)\left(1 + \frac{\Lambda}{3}a^2\right) + \frac{2ma^2}{r}
  \right].
  \label{eq:Dr}
\end{eqnarray}
Two constants of motion of the light ray, the energy $E$ and
the angular momentum $L$, are given by
\begin{eqnarray}
 E &=&
  A(r)\frac{dt}{d\lambda} - C(r)\frac{d\phi}{d\lambda},\\
 L &=&
  C(r)\frac{dt}{d\lambda} + D(r)\frac{d\phi}{d\lambda},
\end{eqnarray}
where $\lambda$ is the affine parameter. Solving for
$dt/d\lambda$ and $d\phi/d\lambda$, we obtain two relations:
\begin{eqnarray}
 \frac{dt}{d\lambda}
  &=&
  \frac{ED(r) + LC(r)}{A(r)D(r) + C^2(r)},
  \label{eq:dtdl}\\
 \frac{d\phi}{d\lambda}
  &=&
  \frac{LA(r) - EC(r)}{A(r)D(r) + C^2(r)}.
  \label{eq:dpdl}
\end{eqnarray}
From the null condition $ds^2 = 0$ and Eqs.
(\ref{eq:kerr-deSittermetric2}), (\ref{eq:dtdl}), and (\ref{eq:dpdl}),
the geodesic equation of the light ray can be expressed as
\begin{eqnarray}
 \left(\frac{dr}{d\phi}\right)^2
  =
  \frac{A(r)D(r) + C^2(r)}{B(r)[bA(r) - C(r)]^2}
  [- b^2A(r) + 2bC(r) + D(r)],
  \label{eq:geodesiceq1}
\end{eqnarray}
where $b$ is the impact parameter defined as
\begin{eqnarray}
 b \equiv \frac{L}{E}.
  \label{eq:impact}
\end{eqnarray}
Using Eqs. (\ref{eq:Ar}), (\ref{eq:Br}), (\ref{eq:Cr}),
(\ref{eq:Dr}), and (\ref{eq:geodesiceq1}),
the first-order differential equation of the light ray becomes
\begin{eqnarray}
 \left(\frac{dr}{d\phi}\right)^2
  &=&
  \frac{\displaystyle{
  \left[
   \left(1 - \frac{\Lambda}{3}r^2\right)(r^2 + a^2) - 2mr
    \right]^2}}
  {\displaystyle{r^2
\left(1 + \frac{\Lambda}{3}a^2\right)^{2}
\left[
       b - \frac{2m}{r}(b - a) - \frac{\Lambda}{3}(b - a)(r^2 + a^2)
      \right]^2}}
  \nonumber\\
 & &\quad\times
  \left[
   r^2 + a^2 - b^2 + \frac{2m}{r}(b - a)^2
   + \frac{\Lambda}{3}(b - a)^2(r^2 + a^2)
  \right].
  \label{eq:geodesiceq2}
\end{eqnarray}
Putting $u = 1/r$, Eq. (\ref{eq:geodesiceq2}) is rewritten as
\footnote{
We mention that the term $\left(1 + \frac{\Lambda}{3}a^2\right)^{2}$
in the denominator is missing in Eq. (4) of \cite{sultana2013}
however, their result is not affected by this missing term
because they expanded Eq. (4) up to the second order in $m$, $a$,
and $\Lambda$, whereas $\Lambda a^2$ corresponds to the third order.}
\begin{eqnarray}
 \left(\frac{du}{d\phi}\right)^2
  &=&
  \frac{\displaystyle{\left[
	 \left(1 - \frac{\Lambda}{3u^2}\right)
	 (1 + a^2 u^2) - 2mu
	\right]^2}}
  {\displaystyle{
  \left(1 + \frac{\Lambda}{3}a^2\right)^{2}
  \left[
    b - 2mu(b - a) - \frac{\Lambda}{3u^2}(b - a)(1 + a^2 u^2)
      \right]^2}}
  \nonumber\\
 & &\quad\times
  \left[
   1 + (a^2 - b^2)u^2 + 2mu^3(b - a)^2
   + \frac{\Lambda}{3}(b - a)^2(1 + a^2 u^2)
	  \right].
  \label{eq:geodesiceq3}
\end{eqnarray}
Expanding Eq. (\ref{eq:geodesiceq3}) up to the order
${\cal O}(m^2, a^2, am, m\Lambda, a\Lambda, \Lambda^2)$,
we have
\begin{eqnarray}
 \left(\frac{du}{d\phi}\right)^2
  &=&
  \frac{1}{b^2} + \frac{\Lambda}{3} - u^2 + 2mu^3
  - 2a^2 u^2 + \frac{3a^2 u^2}{b^2} - \frac{4amu}{b^3}
  + \frac{2\Lambda a}{3b^3 u^2}
  + {\cal O}(\varepsilon^3).
  \label{eq:geodesiceq4}
\end{eqnarray}
Note that for the sake of simplicity, we introduced the notation
for the small expansion parameters $m$, $a$ and $\Lambda$ as
\begin{eqnarray}
 \varepsilon \sim m \sim a \sim \Lambda,\quad
  \varepsilon \ll 1,
\end{eqnarray}
then ${\cal O}(\varepsilon^3)$ in Eq. (\ref{eq:geodesiceq4})
denotes combinations of these three parameters.
Henceforth we use the same notation to represent the order of the
approximation and residual terms.

It is instructive to discuss how to choose a zeroth-order solution
$u_0$ of the light trajectory $u$. If $m = 0$ and $a = 0$,
then Eq. (\ref{eq:geodesiceq4}) reduces to the null geodesic
equation in de Sitter spacetime
\begin{eqnarray}
 \left(\frac{du}{d\phi}\right)^2
  =
  \frac{1}{b^2} + \frac{\Lambda}{3} - u^2,
  \label{eq:geodesiceq5}
\end{eqnarray}
which can be also derived immediately from Eqs. (\ref{eq:geodesiceq2})
and (\ref{eq:geodesiceq3}).
Because we assume a nonzero cosmological constant
$\Lambda$ {\it a priori}, we cannot take $\Lambda$ to be zero;
in fact the action
\begin{eqnarray}
 {\cal S} = \int \left[\frac{c^4}{16\pi G}(R - 2\Lambda)
		  + {\cal L}_{M}\right]\sqrt{-g}d^4 x,
\end{eqnarray}
and the field equation
\begin{eqnarray}
 R_{\mu\nu} - \frac{1}{2}g_{\mu\nu}R + \Lambda g_{\mu\nu}
  = \frac{8\pi G}{c^4} T_{\mu\nu},
\end{eqnarray}
include the cosmological constant $\Lambda$ explicitly where
$g = \det (g_{\mu\nu})$, ${\cal L}_{M}$ denotes the Lagrangian
for the matter field; $R_{\mu\nu}$ and $R$ are the Ricci tensor
and Ricci scalar, respectively; and $T_{\mu\nu}$ is the
energy-momentum tensor. Because $m$ and $a$ are the integration
or arbitrary constants in the Kerr--de Sitter solution, 
it is possible to put $m = 0$ and $a = 0$.
Therefore Eq. (\ref{eq:geodesiceq5}) cannot be reduced to
the null geodesic equation in Minkowski spacetime and the
zeroth-order solution of the $u$ of the light ray should be
taken as the form
\begin{eqnarray}
 u_0 = \frac{\sin\phi}{B},\quad
  \frac{1}{B^2} \equiv \frac{1}{b^2} + \frac{\Lambda}{3}
  \label{eq:u0}
\end{eqnarray}
instead of $u_0 = \sin\phi/b$. Note that Eq. (\ref{eq:u0}) should be
evaluated in de Sitter spacetime
\begin{eqnarray}
 ds^2 = - \left(1 - \frac{\Lambda}{3}r^2\right)dt^2 +
  \left(1 - \frac{\Lambda}{3}r^2\right)^{-1}dr^2 + r^2d\phi^2,
  \label{eq:deSitter}
\end{eqnarray}
instead of in the Minkowski spacetime. The choice of the zeroth-order
solution is key to obtaining the measurable angle $\psi$;
see section \ref{sec:static} below.

Let us obtain the equation of the light trajectory in accordance
with the standard perturbation scheme. We take the solution
$u = u(\phi)$ of the light trajectory as
\begin{eqnarray}
 u = \frac{\sin \phi}{B} + \delta u_1 + \delta u_2,
  \label{eq:trajectory1}
\end{eqnarray}
where $\delta u_1$ and $\delta u_2$ are the first order
${\cal O}(\varepsilon)$ and second order ${\cal O}(\varepsilon^2)$
corrections to the zeroth-order solution $u_0 = \sin\phi/B$,
respectively. Inserting Eq. (\ref{eq:trajectory1}) into
Eq. (\ref{eq:geodesiceq4}), then expanding, integrating  and collecting
the same order terms, the equation of the light trajectory in Kerr--de
Sitter spacetime is given up to second-order ${\cal O}(\varepsilon^2)$
by 
\begin{eqnarray}
\frac{1}{r}
  &=& \frac{\sin\phi}{B}
  + \frac{m}{2B^2}(3 + \cos 2\phi)\nonumber\\
 & &
  + \frac{1}{16 B^3 b^3}
  \left\{
  b^3\left(3 m^2 + 2 a^2 \right)(3\sin\phi - \sin 3\phi)
  +
  4b\left[7 b^2 m^2 + 6 a^2 (B^2 - b^2)\right]\sin\phi
  \right.
  \nonumber\\
 & &+\left.
      \left(
	     30 b^3 m^2 - 12 a^2 b^3 + 12 B^2 a^2 b  
      \right) \left(\pi - 2\phi\right) \cos\phi - 32 B^3 a m
	  \right\}
  \nonumber\\
 & &-
  \frac{\Lambda B^3 a\left(2\sin^2\phi - 1\right)}
  {3 b^3 \sin\phi}
 + {\cal O}(\varepsilon^3),
  \label{eq:trajectory2}
\end{eqnarray}
where the integration constants of $\delta u_1$ and $\delta u_2$
are chosen so as to maximize $u$ (or to minimize $r$) at
$\phi = \pi/2$
\begin{eqnarray}
 \left.
  \frac{d\delta u_1}{d\phi}
  \right|_{\phi = \pi/2} = 0,\quad
  \left.
  \frac{d\delta u_2}{d\phi}
     \right|_{\phi = \pi/2} = 0.
  \label{eq:int_const}
\end{eqnarray}

Eq. (\ref{eq:trajectory2}) contains two constants $B$ and $b$,
and leads to a complicated expression for the measurable angle
$\psi$. To avoid the complex expression for $\psi$,
we expand $B$ in $\Lambda$ and express $B$ by $b$, obtaining
\begin{eqnarray}
\frac{1}{r}
 &=& \left(
      \frac{1}{b}
      + \frac{b\Lambda}{6}
      - \frac{b^3\Lambda^2}{72}
      \right)\sin\phi
 + \frac{m}{2}(3 + \cos 2\phi)
 \left(\frac{1}{b^2} + \frac{\Lambda}{3}\right)\nonumber\\
 & &
  + \frac{1}{16 b^3}
  \left\{
   m^2\left[37\sin\phi + 30(\pi - 2\phi) \cos\phi - 3\sin 3\phi
       \right]
       + 8a^2\sin^3\phi- 32 a m
  \right\}\nonumber\\
  & &-
  \frac{\Lambda a\left(2\sin^2\phi - 1\right)}
  {3 \sin\phi}
 + {\cal O}(\varepsilon^3).
  \label{eq:trajectory3}
\end{eqnarray}
Note that if we use the approximate solution of
Eq. (\ref{eq:geodesiceq4}) as
\footnote{This solution is used by \cite{sultana2013}.}
\begin{eqnarray}
 u = \frac{\sin\phi}{b} + \delta u_1 + \delta u_2,
  \label{eq:trajectory5}
\end{eqnarray}
the following terms in Eq. (\ref{eq:trajectory3}) disappear:
\begin{eqnarray}
 \left(\frac{b\Lambda}{6} - \frac{b^3\Lambda^2}{72}\right)\sin\phi,
  \quad
  \frac{m\Lambda}{6}(3 + \cos 2\phi).
  \label{eq:trajectory6}
\end{eqnarray}
The existence of the above terms in Eq. (\ref{eq:trajectory3})
reflects the fact that the background spacetime is de Sitter spacetime
instead of Minkowski spacetime.

Before concluding this section, it is noteworthy that
unlike Schwarzschild--de Sitter spacetime, the trajectory equation
of the light ray in Kerr--de Sitter spacetime depends on
the cosmological constant $\Lambda$; from the condition 
\begin{eqnarray}
 \left.\frac{du}{d\phi}\right|_{u = u_0} = 0,\quad
  u_0 = \frac{1}{r_0},
\end{eqnarray}
and Eq. (\ref{eq:geodesiceq4}), we have following relation:
\begin{eqnarray}
 \frac{1}{B^2} = \frac{1}{b^2} + \frac{\Lambda}{3}
  =
  \frac{1}{r_0^2} - \frac{2m}{r_0^3} + \frac{2a^2}{r_0^2}
  - \frac{3a^2}{b^2r_0^2} + \frac{4am}{b^3r_0}
  - \frac{2\Lambda a r_0^2}{3b^3}
  + {\cal O}(\varepsilon^3),
  \label{eq:closest}
\end{eqnarray}
in which $r_0$ is the radial coordinate value of the light ray
at the point of closest approach (in our case $\phi = \pi/2$),
and $r_0$ can be obtained by the observation in principle
as the circumference radius $\ell_0 = 2\pi r_0$.
It is also possible to obtain a similar relation from Eqs.
(\ref{eq:geodesiceq3}) and (\ref{eq:geodesiceq4}); but the expression
becomes more complicated.
Eq. (\ref{eq:closest}) means that unlike the Schwarzschild--de Sitter
case, $B$ cannot be expressed only by $r_0$, $m$, and $a$;
$\Lambda$ and $b$ too are required.
As a result, the trajectory equation of the light ray
in Kerr--de Sitter spacetime depends on the cosmological
constant $\Lambda$ and $b$; whereas the equation of the light trajectory
in Schwarzschild--de Sitter spacetime is independent of
$\Lambda$ and $b$; in fact setting $a = 0$ in Eqs.
(\ref{eq:geodesiceq2}) and (\ref{eq:geodesiceq4}) yields
\begin{eqnarray}
  \left(\frac{du}{d\phi}\right)^2
  =
  \frac{1}{b^2} + \frac{\Lambda}{3} - u^2 + 2mu^3,
  \label{eq:SdS}
\end{eqnarray}
thus using $u_0 = 1/r_0$, we obtain
\begin{eqnarray}
 \frac{1}{B^2_{\rm SdS}} = \frac{1}{b^2} + \frac{\Lambda}{3}
  =
  \frac{1}{r_0^2} - \frac{2m}{r_0^3}.
  \label{eq:closest_SdS}
\end{eqnarray}
Eq. (\ref{eq:closest_SdS}) shows that the constant $B_{\rm SdS}$ in
Schwarzschild--de Sitter spacetime can be determined without
knowing $\Lambda$ and $b$.
%%%%%%%%%%%%%%%%%%%%%%%%%%%%%%%%%%%%%%%%%%%%%%%%%%%%%%%%%%%%%%%%%%
\section{General Relativistic Aberration Equation
\label{sec:aberration}}
The general relativistic aberration equation is given by
\cite{lebedevlake2013}; also see
\cite{pechenick_etal1983,lebedevlake2016}:
\begin{eqnarray}
 \cos\psi =
  \frac{g_{\mu\nu}k^{\mu}w^{\nu}}
  {(g_{\mu\nu}u^{\mu}k^{\nu})(g_{\mu\nu}u^{\mu}w^{\nu})} + 1,
  \label{eq:aberration1}
\end{eqnarray}
where $k^{\mu}$ is the 4-momentum of the light ray $\Gamma_k$ which
we investigate, $w^{\mu}$ is the 4-momentum of the radial null
geodesic $\Gamma_w$ connecting the center $O$ and the position of
observer $P$, $u^{\mu} = dx^{\mu}/d\tau$ is the 4-velocity of
the observer ($\tau$ is the proper time of the observer),
and $\psi$ is the angle between the two vectors $k^{\mu}$ and $w^{\mu}$
at the position of observer $P$. The details of the derivation of
Eq. (\ref{eq:aberration1}) are described in section V B of
\cite{lebedevlake2013}. Because Eq. (\ref{eq:aberration1}) includes
the 4-velocity of the observer $u^{\mu}$, it enables us to calculate
the influence of the motion of the observer on the measurable
angle $\psi$. 
\begin{figure}[htbp]
\begin{center}
 \includegraphics[scale=0.3,clip]{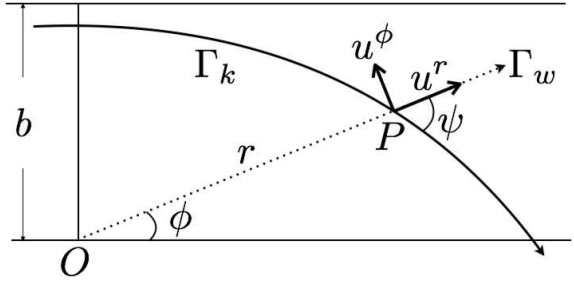}
 \caption{Schematic diagram of light trajectory.
 $\Gamma_k$ (bold line) is the trajectory of the light ray
 we investigate, $\Gamma_w$ (dotted line) is the radial null
 geodesic connecting the center $O$ and the position of observer $P$
 and the measurable angle $\psi$ is the intersection angle between
 $\Gamma_k$ and $\Gamma_w$ at $P$. Two bold vectors $u^r$ and $u^{\phi}$
 at $P$ indicate the directions of the $r$ (radial) and $\phi$
 (transverse) components of the 4-velocity $u^{\mu}$.
 The direction of time component
 $u^t$ is perpendicular to this schematic plane.
 \label{fig:arakida-fig1}}
\end{center}
\end{figure}
See FIG. \ref{fig:arakida-fig1} for the schematic diagram of
the light trajectory.

Eq. (\ref{eq:aberration1}) can be written as the tangent formula
(see Eq. (22) in \cite{arakida2018b}) which gives the same approximate
solution of the measurable angle $\psi$ as with Eq.
(\ref{eq:aberration1}) but requires tedious and lengthy
calculations. Hence, we will present the results obtained from
Eq. (\ref{eq:aberration1}).

%%%%%%%%%%%%%%%%%%%%%%%%%%%%%%%%%%%%%%%%%%%%%%%%%%%%%%%%%%%%%
\section{Measurable Angle in Kerr--de Sitter Spacetime
\label{sec:angle}}
Henceforth, in accordance with the procedure described in our previous
paper \cite{arakida2018b}, we calculate the measurable angle $\psi$ at
the position of observer $P$. From now on, only important equations
are summarized; see section IV in \cite{arakida2018b} for the details
of the derivation.

Because we are working in the equatorial plane $\theta = \pi/2, d\theta = 0$
as the orbital plane of the light ray, the components of the tangent
vectors $k^{\mu}$ and $w^{\mu}$ are
\begin{eqnarray}
 k^{\mu} &=& (k^t, k^r, 0, k^{\phi}),\\
 w^{\mu} &=& (w^t, w^r, 0, 0).
\end{eqnarray}
As $k^{\mu}$ and $w^{\mu}$ are null vectors, from the null
condition $g_{\mu\nu}k^{\mu}k^{\nu} = 0$ and
$g_{\mu\nu}w^{\mu}w^{\nu} = 0$, $k^t$ and $w^t$ are expressed
in terms of $k^r$, $k^{\phi}$ and $w^r$, respectively as
\begin{eqnarray}
 k^t &=& \frac{C(r)k^{\phi} +
  \sqrt{[C(r)k^{\phi}]^2 + A(r)[B(r)(k^r)^2 + D(r)(k^{\phi})^2]}}
  {A(r)},
  \label{eq:kt}\\
 w^t &=& \sqrt{\frac{B(r)}{A(r)}}w^r,
  \label{eq:wt}
\end{eqnarray}
where we chose the sign of $k^t$ and $w^t$ to be positive.
The inner product $g_{\mu\nu}k^{\mu}w^{\nu}$ is given 
in terms of $A(r)$, $B(r)$, $C(r)$, and $D(r)$ by
\begin{eqnarray}
 g_{\mu\nu}k^{\mu}w^{\nu}
 &=&
  \left\{
  - 
  \sqrt{\frac{B(r)}{A(r)}}
  \frac{A(r)D(r) + C^2(r)}{bA(r) - C(r)}
  \right.\nonumber\\
 & &\left.
  + \frac{\sqrt{B(r)[A(r)D(r) + C^2(r)][-b^2A(r) + 2bC(r) + D(r)]}}
  {bA(r) - C(r)}\right\}k^{\phi}w^r.
  \label{eq:kw}
\end{eqnarray}
%%%%%%%%%%%%%%%%%%%%%%%%%%%%%%%%%%%%%%%%%%%%%%%%%%%%%%%%%%%%%%%%%%%%
\subsection{Measurable Angle by Static Observer
\label{sec:static}}
In the case of the static observer, the component of
the 4-velocity of the observer $u^{\mu}$ becomes
\begin{eqnarray}
 u^{\mu} = (u^t, 0, 0, 0),
\end{eqnarray}
and the condition for the time-like observer
$g_{\mu\nu}u^{\mu}u^{\nu} = -1$ gives $u^t$ as
\begin{eqnarray}
 u^t = \frac{1}{\sqrt{A(r)}},
  \label{eq:ut_sta}
\end{eqnarray}
where we take $u^t$ to be positive. The inner products
$g_{\mu\nu}u^{\mu}k^{\nu}$ and $g_{\mu\nu}u^{\mu}w^{\nu}$
become
\begin{eqnarray}
 g_{\mu\nu}u^{\mu}k^{\nu}
 &=&
  -\frac{1}{\sqrt{A(r)}}
  \frac{A(r)D(r) + C^2(r)}{bA(r) - C(r)}k^{\phi},
  \label{eq:uk_sta}\\
 g_{\mu\nu}u^{\mu}w^{\nu}
  &=& -\sqrt{B(r)}w^r,
  \label{eq:uw_sta}
\end{eqnarray}
Inserting Eqs. (\ref{eq:kw}), (\ref{eq:uk_sta}), and (\ref{eq:uw_sta})
into Eq. (\ref{eq:aberration1}), we have
\begin{eqnarray}
 \cos\psi_{\rm static}
  = \sqrt{\frac{A(r)[-b^2A(r) + 2bC(r) + D(r)]}{A(r)D(r) + C^2(r)}}.
  \label{eq:cos_aberration_sta1}
\end{eqnarray}
Further, substituting Eqs. (\ref{eq:Ar}), (\ref{eq:Br}), (\ref{eq:Cr}),
(\ref{eq:Dr}) and (\ref{eq:trajectory3})
into Eq. (\ref{eq:cos_aberration_sta1}), and expanding up to the
order ${\cal O}(\varepsilon^2)$, $\psi_{\rm static}$  
for the range $0 \le \psi \le \pi/2$ is given by
\begin{eqnarray}
 \psi_{\rm static}
  &=& \phi + \frac{2m}{b}\cos\phi
  \nonumber\\
  & &+ \frac{1}{8b^2}
  \left\{
   m^2\left[15(\pi - 2\phi) - \sin 2\phi\right]
   - 16ma \cos \phi
		  \right\}
  \nonumber\\
 & &- \frac{\Lambda b^2}{6}\cot \phi
  + \frac{\Lambda b}{3}\cos\phi
  \left[m(1 + \csc^2\phi) + 2a\csc\phi\right]
  - \frac{\Lambda^2 b^4}{288}\csc^4\phi\sin 4\phi
  \nonumber\\
 & &+ {\cal O}(\varepsilon^3).
  \label{eq:psi_sta1}
\end{eqnarray}
and for the range $\pi/2 \le \phi \le \pi$
\begin{eqnarray}
 \psi_{\rm static}
  &=& \pi - \phi- \frac{2m}{b}\cos\phi
  \nonumber\\
  & &- \frac{1}{8b^2}
  \left\{
   m^2\left[15(\pi - 2\phi) - \sin 2\phi\right]
   - 16ma \cos \phi
		  \right\}
  \nonumber\\
 & &+ \frac{\Lambda b^2}{6}\cot\phi
  - \frac{\Lambda b}{3}\cos\phi
  \left[m(1 + \csc^2\phi) + 2a\csc\phi\right]
  + \frac{\Lambda^2 b^4}{288}\csc^4\phi\sin 4\phi
  \nonumber\\
 & &+ {\cal O}(\varepsilon^3).
  \label{eq:psi_sta2}
\end{eqnarray}
As in \cite{arakida2018b}, we divided the expression for
$\psi_{\rm static}$ into two cases, Eqs. (\ref{eq:psi_sta2}) and
(\ref{eq:psi_sta1}). The purpose of this was to utilize trigonometric
identities such as
$\sqrt{1 - \sin^2\phi} = \cos\phi$ for $0 \le \phi \le \pi/2$ and
$\sqrt{1 - \sin^2\phi} = -\cos\phi$ for $\pi/2 \le \phi \le \pi$.
Henceforth we adopt a similar procedure when calculating angle
measured by the observer in radial motion, $\psi_{\rm radial}$ and
in transverse motion $\psi_{\rm transverse}$ below. Although this
procedure may not be necessary for computing $\psi_{\rm static}$
and $\psi_{\rm radial}$, it is required when
computing $\psi_{\rm transverse}$; see Eq. (\ref{eq:psi_circ1})
and observe the case for $\phi \rightarrow \pi$.

The first and second lines in Eqs. (\ref{eq:psi_sta1}) and
(\ref{eq:psi_sta2}) are in agreement with the measurable angle of
the static observer in Kerr spacetime derived in \cite{arakida2018b},
and the third lines in Eqs. (\ref{eq:psi_sta1}) and (\ref{eq:psi_sta2})
are due to the influence of the cosmological constant $\Lambda$.

Here, let us estimate how the cosmological constant $\Lambda$
contributes to the measurable angle of the light ray.
We assume that the observer is located within the range
$0 \le \phi \le \pi/2$ and as the lens object, we adopt the
typical galaxy with mass $M_{\rm gal}$, radius $R_{\rm gal}$,
and angular momentum $J_{\rm gal}$; the impact parameter $b$
is comparable with $R_{\rm gal}$ (see TABLE \ref{tab:arakida-table1}).
\begin{table}[htbp]
 \caption{\label{tab:arakida-table1} Numerical values.
 We use the following numerical values in this paper.
 As the value of the total angular momentum $J_{\rm gal}$,
 we adopt that of our Galaxy
 $J_{\rm gal} \simeq 1.0 \times 10^{67}$ kg m$^{2}$/s from
 \cite{karachentsev1987}.}
\begin{ruledtabular}
\begin{tabular}{lcr}
\textrm{Name}&
\textrm{Symbol}&
\textrm{Value}\\
 \colrule
 Mass of the Galaxy & $M_{\rm gal} = 10^{12}M_{\odot}$ &
	 $2.0 \times 10^{42} ~{\rm kg}$\\
                    & $m = GM_{\rm gal}/c^2$ &
	 $1.5 \times 10^{15}$ m\\
 Impact Parameter & $b = R_{\rm gal} = 26 ~{\rm kly}$ &
	 $2.5 \times 10^{20}$ m\\
 Angular Momentum of the Galaxy \cite{karachentsev1987}
 & $J = J_{\rm gal}$ & $1.0 \times 10^{67}$ kg m$^2$/s\\
 Spin Parameter & $a = J/(M_{\rm gal}c)$ & $1.7 \times 10^{16}$ m\\ 
 Cosmological Constant & $\Lambda$ & $10^{-52}$ m$^{-2}$ \\
 Hubble Constant & $H_0 = c\sqrt{\Lambda/3}$ &
	 $1.73 \times 10^{-18} ~{\rm s}^{-1}$\\
 Distance from Lens Object & $D = 1.0 ~{\rm Gly}$ &
	 $9.4 \times 10^{24} ~{\rm m}$\\
 Recession Velocity & $v_H = H_0 D$ & $1.6 \times 10^{7} ~{\rm m/s}$ \\
 Radial Velocity & $v^r = v_H/c$ & $0.05$\\
 Transverse Velocity & $bv^{\phi} = v_H/c$ & $0.05$\\
\end{tabular}
\end{ruledtabular}
\end{table}
Because the Kerr contributions appearing in Eqs. (\ref{eq:psi_sta1}) and
(\ref{eq:psi_sta2}) are examined in our previous paper
\cite{arakida2018b}, we extract the terms concerning
the cosmological constant $\Lambda$ from Eq. (\ref{eq:psi_sta1})
and put
\begin{eqnarray}
 \psi_{\rm static}(\phi; \Lambda, b) &=&
  - \frac{\Lambda b^2}{6}\cot \phi
  - \frac{\Lambda^2 b^4}{288}\csc^4\phi\sin 4\phi,
  \label{eq:psi_sta_part1}\\
 \psi_{\rm static}(\phi: m, a, \Lambda ,b)
  &=& \frac{\Lambda b}{3}\cos\phi
  \left[m(1 + \csc^2\phi) + 2a\csc\phi\right],
  \label{eq:psi_sta_part2}
\end{eqnarray}

To compare the contribution of the cosmological constant
$\Lambda$ with the result of the Kerr case, we compute the following
terms of the total deflection angle $\alpha$ in Kerr spacetime
(see e.g., \cite{arakida2018b}) using the values summarized
in TABLE \ref{tab:arakida-table1}:
\begin{eqnarray}
 \frac{4m}{b} &\approx& 2.4 \times 10^{-5} ~{\rm rad},
  \label{eq:def_angle1}\\
 \frac{15\pi m^2}{4b^2} &\approx& 4.2 \times 10^{-10} ~{\rm rad},
  \label{eq:def_angle2}\\
 - \frac{4ma}{b^2} &\approx& \mp 1.6 \times 10^{-9} ~{\rm rad}\quad
  \mbox{for}\quad \pm a.
  \label{eq:def_angle3}
\end{eqnarray}

\begin{figure}[htbp]
\begin{center}
 \includegraphics[scale=1.0,clip]{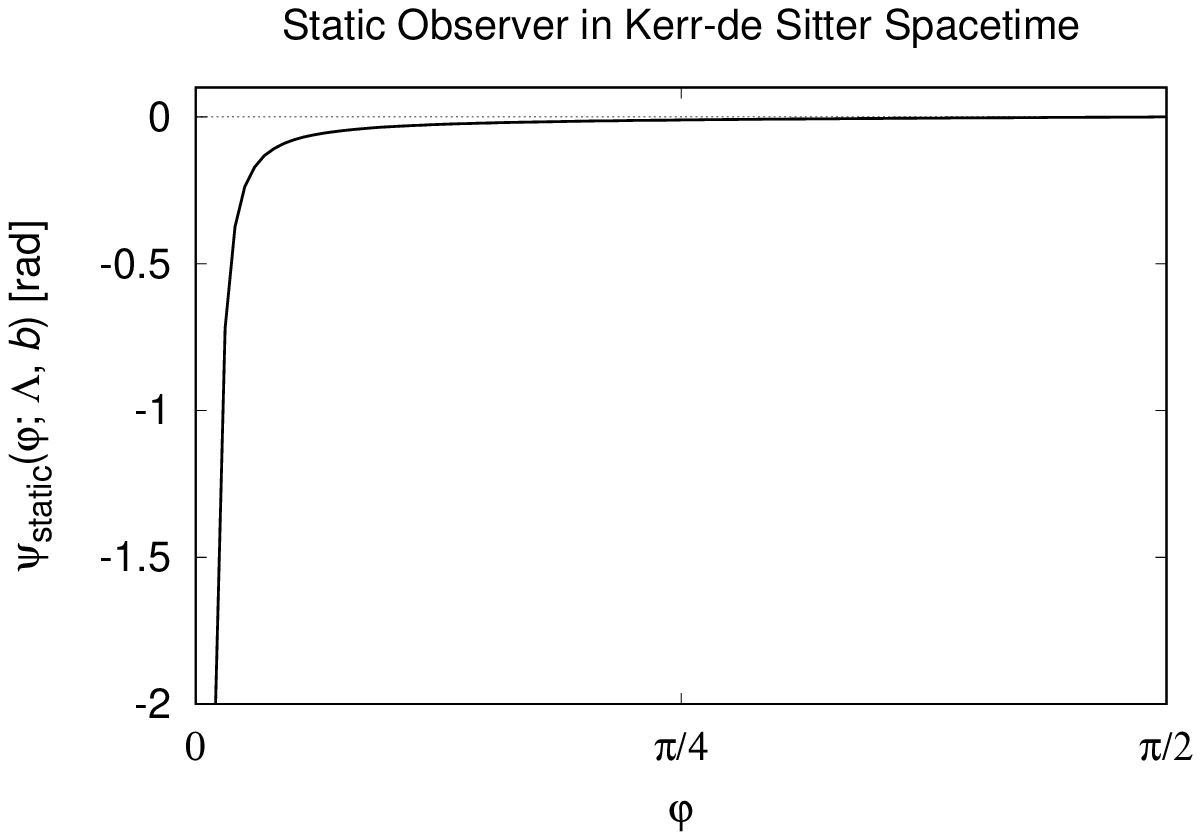}
 \caption{$\phi$ dependence of Eq. (\ref{eq:psi_sta_part1}).
 \label{fig:arakida-fig2}}
\end{center}
\end{figure}

\begin{figure}[htbp]
\begin{center}
 \includegraphics[scale=1.0,clip]{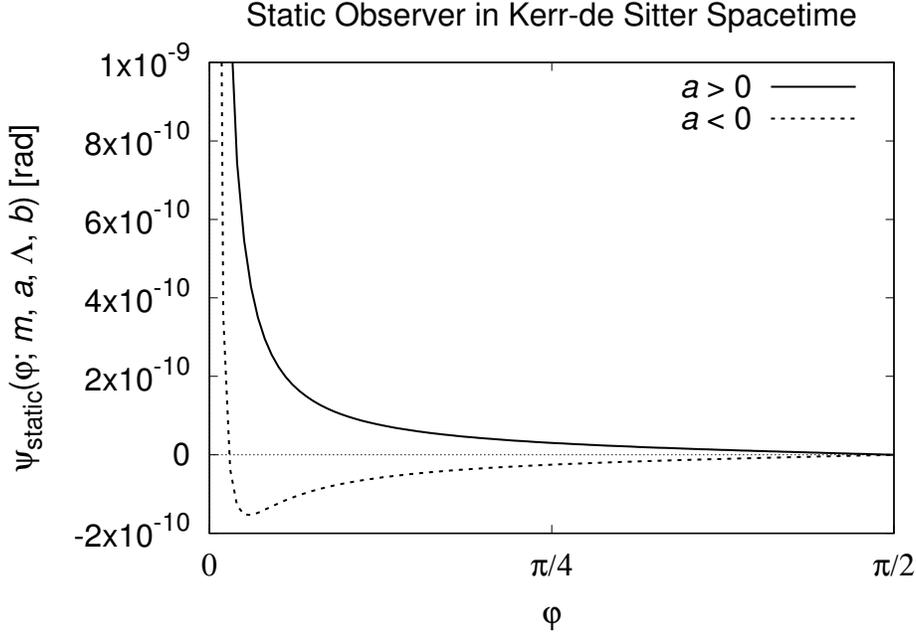}
 \caption{$\phi$ dependence of Eq. (\ref{eq:psi_sta_part2}).
 \label{fig:arakida-fig3}}
\end{center}
\end{figure}

FIGs. \ref{fig:arakida-fig2} and \ref{fig:arakida-fig3}
show the $\phi$ dependences of Eqs. (\ref{eq:psi_sta_part1}) and
(\ref{eq:psi_sta_part2}), respectively.
From FIG. \ref{fig:arakida-fig2}, Eq. (\ref{eq:psi_sta_part1})
is a monotonic function of $\phi$ and increases rapidly, 
diverging to negative infinity as $\phi$ approaches 0.
This property is due to the existence of the de Sitter horizon at
$r_{\rm dS} \approx \sqrt{3/\Lambda}$.
The order ${\cal O}(\Lambda b, \Lambda^2 b^2)$ terms in
Eq. (\ref{eq:psi_sta_part1}) take a negative value which
diminishes the measurable angle $\psi$.

Eq. (\ref{eq:psi_sta_part2}) contains the order
${\cal O}(\Lambda bm, \Lambda ba)$ terms and its magnitude is
${\cal O}(10^{-10})$ which is almost comparable to the second order
deflection angle, Eq. (\ref{eq:def_angle2}).
In accordance with the sign of the spin parameter $a$,
Eq. (\ref{eq:psi_sta_part2}) takes a different sign;
for $a > 0$, Eq. (\ref{eq:psi_sta_part2}) is positive
and vice versa. However, regardless of the sign of the
spin parameter $a$, Eq. (\ref{eq:psi_sta_part2}) diverges to
positive infinity as $\phi$ approaches $0$.

Before concluding this section, we note that 
if Eq. (\ref{eq:trajectory5}) (see also Eq. (\ref{eq:trajectory6}))
is adopted instead of Eqs.
(\ref{eq:trajectory1}) and (\ref{eq:trajectory3}), the measurable angle
$\tilde{\psi}_{\rm static}$ for the range $0 \le \phi \le \pi/2$ becomes
\begin{eqnarray}
 \tilde{\psi}_{\rm static}
  &=& \phi + \frac{2m}{b}\cos\phi
  \nonumber\\
  & &+ \frac{1}{8b^2}
  \left\{
   m^2\left[15(\pi - 2\phi) - \sin 2\phi\right]
   - 16ma \cos \phi
		  \right\}
  \nonumber\\
 & &- \frac{\Lambda b^2}{6}\csc \phi\sec\phi
  + \frac{\Lambda b}{3}
  \left[
   m(\cot\phi\csc\phi - \sec\phi) + 2a\cot\phi
	 \right]
  \nonumber\\
  & &- \frac{\Lambda^2 b^4}{9}\cot 2\phi\csc^2 2\phi
  + {\cal O}(\varepsilon^3),
  \label{eq:psi_sta3}
\end{eqnarray}
where the first and second lines are also in agreement with
the measurable angle of the static observer in the Kerr spacetime
as derived in \cite{arakida2018b}.

Comparing Eqs. (\ref{eq:psi_sta1}) and (\ref{eq:psi_sta3}),
we find the following: first, in spite of the different correction
terms due to the cosmological constant $\Lambda$,
the measurable angles $\psi_{\rm static}$ and
$\tilde{\psi}_{\rm static}$ take a large value rapidly
and diverge to negative infinity when $\phi$ approaches 0.
This property is related to the existence of the de Sitter horizon.
Second, when $\phi \rightarrow \pi/2$, Eq. (\ref{eq:psi_sta1}) leads
to the result $\psi_{\rm static} \rightarrow \pi/2$,
which is consistent with the initial condition in our case,
Eq. (\ref{eq:int_const}). Note that at $\phi = \pi/2$,
the two null geodesics $k^{\mu}$ and $w^{\mu}$ are orthogonal.
However, Eq. (\ref{eq:psi_sta3}) diverges, and
$\tilde{\psi}_{\rm static} \rightarrow \infty$, which contradicts
the initial condition Eq. (\ref{eq:int_const}).
Therefore, Eq. (\ref{eq:u0}) should be employed as the zeroth-order
solution of $u$; as a consequence Eq. (\ref{eq:trajectory2}) or at least
Eq. (\ref{eq:trajectory3}) should be used as the trajectory equation
of the light ray when investigating light bending in Kerr--de
Sitter spacetime. The same holds for Schwarzschild--de Sitter
spacetime. Until now, how a zeroth-order solution $u_0$ is to be
chosen was not considered carefully, the above indication is one of
the important suggestions in this paper.

%%%%%%%%%%%%%%%%%%%%%%%%%%%%%%%%%%%%%%%%%%%%%%%%%%%%%%%%%%%%%%%
\subsection{Measurable Angle by Observer in Radial Motion}
The component of the 4-velocity $u^{\mu}$ of the radially
moving observer is
\begin{eqnarray}
 u^{\mu} = (u^t, u^r, 0, 0),
\end{eqnarray}
and from the condition $g_{\mu\nu}u^{\mu}u^{\nu} = -1$,
$u^t$ can be expressed in terms of $u^r$ as
\begin{eqnarray}
 u^t = \sqrt{\frac{B(r)(u^r)^2 + 1}{A(r)}},
  \label{eq:ut_rad}
\end{eqnarray}
where $u^t$ is taken to be positive.
The inner products $g_{\mu\nu}u^{\mu}k^{\nu}$ and
$g_{\mu\nu}u^{\mu}w^{\nu}$ are given by
\begin{eqnarray}
 g_{\mu\nu}u^{\mu}k^{\nu}
 &=&
  \left\{
  -\sqrt{\frac{B(r)(u^r)^2 + 1}{A(r)}}
  \frac{A(r)D(r) + C^2(r)}{bA(r) - C(r)}
		    \right.\nonumber\\
 & &+\left.
  u^r
  \frac{\sqrt{B(r)[A(r)D(r) + C^2(r)][-b^2A(r) + 2bC(r) + D(r)]}}
  {bA(r) - C(r)}
  \right\}k^{\phi},
 \label{eq:uk_rad}\\
 g_{\mu\nu}u^{\mu}w^{\nu}
  &=& \left\{
     - \sqrt{B(r)\left[B(r)(u^r)^2 + 1\right]} + B(r)u^r
    \right\}w^r.
  \label{eq:uw_rad}
\end{eqnarray}
Here, instead of $u^r$, we introduce the coordinate radial
velocity $v^r$ as
\begin{eqnarray}
 v^r = \frac{dr}{dt} = \frac{dr/d\tau}{dt/d\tau} = \frac{u^r}{u^t},
\label{eq:vr1}
\end{eqnarray}
and substituting Eq. (\ref{eq:ut_rad}) into Eq. (\ref{eq:vr1}),
we find,
\begin{eqnarray}
 u^r = \frac{v^r}{\sqrt{A(r) - B(r)(v^r)^2}}.
  \label{eq:vr2}
\end{eqnarray}
Using Eq. (\ref{eq:vr2}), we rewrite Eqs. (\ref{eq:uk_rad}) and
(\ref{eq:uw_rad}) in terms of $v^r$
\begin{eqnarray}
 g_{\mu\nu}u^{\mu}k^{\nu}
  &=&
  \left\{
  - 
  \frac{A(r)D(r) + C^2(r)}{[bA(r) - C(r)]\sqrt{A(r) - B(r)(v^r)^2}}
    \right.\nonumber\\
 & &+\left.
  v^r
  \frac{\sqrt{B(r)[A(r)D(r) + C^2(r)][-b^2A(r) + 2bC(r) + D(r)]}}
  {[bA(r) - C(r)]\sqrt{A(r) - B(r)(v^r)^2}}
  \right\}k^{\phi},
 \label{eq:uk_rad2}\\
 g_{\mu\nu}u^{\mu}w^{\nu}
  &=&
  - \frac{\sqrt{A(r)B(r)} - B(r)v^r}{\sqrt{A(r) - B(r)(v^r)^2}}
    w^r.
  \label{eq:uw_rad2}
\end{eqnarray}

Here we impose the slow motion approximation for the radial velocity
of the observer, $v^r \ll 1$. Next, following the same procedure
used to obtain Eqs. (\ref{eq:psi_sta1}) and (\ref{eq:psi_sta2}),
we insert Eqs. (\ref{eq:Ar}), (\ref{eq:Br}), (\ref{eq:Cr}),
(\ref{eq:Dr}), (\ref{eq:kw}), (\ref{eq:uk_rad2}) and
(\ref{eq:uw_rad2}) into Eq. (\ref{eq:aberration1}), obtaining
$\psi_{\rm radial}$ for the range $0 \le \phi \le \pi/2$
up to the order ${\cal O}(\varepsilon^2, \varepsilon^2 v^r)$ 
\begin{eqnarray}
 \psi_{\rm radial}
  &=&
  \phi + v^r\sin\phi
  + \frac{2m}{b}(\cos\phi + v^r)
  \nonumber\\
  & &+ \frac{1}{8b^2}
  \left(
   m^2[15(\pi - 2\phi) - \sin 2\phi] - 16am\cos\phi
   \right)\nonumber\\
 & &+ \frac{v^r}{16b^2}\left\{
	m^2\left[
      30 (\pi - 2\phi) \cos\phi + 95\sin\phi - \sin 3\phi
     \right]\right.
	\nonumber\\
    & &\left.-16am(1 + \cos 2\phi) - 2a^2(3\sin\phi - \sin 3\phi)\right\}
     \nonumber\\
  & &- \frac{\Lambda b^2}{6}\cot\phi
  - \frac{\Lambda b^2v^r}{12}(\cos 2\phi - 3)
  + \frac{\Lambda b}{3}\cos\phi
  \left[m(1 + \csc^2\phi) + 2a\csc\phi\right]
  \nonumber\\
  & &+ \frac{\Lambda bv^r}{3}\csc\phi
  \left[
   a + m\csc\phi + (a - 2m \csc\phi)\cos 2\phi 
	 \right]
  \nonumber\\
 & &- \frac{\Lambda^2 b^4}{288}\csc^4\phi\sin 4\phi
  - \frac{\Lambda^2 b^4 v^r}{144}
  (\cos 2\phi - 7)\cot^2\phi \csc\phi
  + {\cal O}(\varepsilon^3, (v^r)^2),
\label{eq:psi_rad1}
\end{eqnarray}
and for the range $\pi/2 \le \phi \le \pi$
\begin{eqnarray}
 \psi_{\rm radial}
  &=&
  \pi - \phi + v^r\sin\phi +
  \frac{2m}{b}\left(-\cos\phi + v^r\right)
  \nonumber\\
  & &- \frac{1}{8b^2}
  \left(
   m^2[15(\pi - 2\phi) - \sin 2\phi] - 16am\cos\phi
     \right)\nonumber\\
 & &
  +\frac{v^r}{16b^2}
  \left\{
   m^2\left[
   30(\pi - 2\phi) \cos\phi + 95\sin\phi - \sin 3\phi
 \right]\right.
  \nonumber\\
 & &\left.- 16am(1 + \cos 2\phi) - 2a^2(3\sin\phi - \sin 3\phi)\right\}
  \nonumber\\
 & &+ \frac{\Lambda b^2}{6}\cot\phi
  - \frac{\Lambda b^2v^r}{12}(\cos 2\phi - 3)
  - \frac{\Lambda b}{3}\cos\phi
  \left[m(1 + \csc^2\phi) + 2a\csc\phi\right]
  \nonumber\\
 & &+ \frac{\Lambda bv^r}{3}\csc\phi
  \left[
   a + m\csc\phi + (a - 2m \csc\phi)\cos 2\phi 
	 \right]
  \nonumber\\
 & &+ \frac{\Lambda^2 b^4}{288}\csc^4\phi\sin 4\phi
  - \frac{\Lambda^2 b^4 v^r}{144}
  (\cos 2\phi - 7)\cot^2\phi \csc\phi
 + {\cal O}(\varepsilon^3, (v^r)^2).
  \label{eq:psi_rad2}
\end{eqnarray}
The first four lines in Eqs. (\ref{eq:psi_rad1}) and (\ref{eq:psi_rad2})
coincide with the measurable angle in Kerr spacetime which has
already been investigated in \cite{arakida2018b}, and the remaining
terms, lines 5 to 7, are the correction
due to the cosmological constant $\Lambda$.

Now we concentrate on investigating the influence of
the cosmological constant $\Lambda$ and the radial velocity
$v^r$. Then we extract the order
${\cal O}(\Lambda b^2 v^r, \Lambda^2 b^4 v^r)$ and
${\cal O}(\Lambda b m v^r, \Lambda b a v^r)$ terms 
from Eq. (\ref{eq:psi_rad1}) except the contributions of
Eqs. (\ref{eq:psi_sta_part1}) and (\ref{eq:psi_sta_part2})
and put
\begin{eqnarray}
 \psi_{\rm radial}(\phi; \Lambda, b, v^r)
  &=& - \frac{\Lambda b^2v^r}{12}(\cos 2\phi - 3)
  - \frac{\Lambda^2 b^4 v^r}{144}
  (\cos 2\phi - 7)\cot^2\phi \csc\phi,
  \label{eq:psi_radial_part1}\\
 \psi_{\rm radial}(\phi; m, a, \Lambda, b, v^r)
  &=&
  \frac{\Lambda bv^r}{3}\csc\phi
  \left[a + m\csc\phi + (a - 2m \csc\phi)\cos 2\phi\right].
  \label{eq:psi_radial_part2}
\end{eqnarray}
Because the background spacetime of Kerr--de Sitter is de Sitter
spacetime, we assume that radial velocity obeys Hubble's law:
\begin{eqnarray}
 v^r \approx v_H = H_0 D,\quad
  H_0 = c\sqrt{\frac{\Lambda}{3}},
  \label{eq:hubble}
\end{eqnarray}
where $D$ is the distance between the lens (central) object $O$
and observer $P$, and we take
$D \approx 1 ~{\rm Gly} \simeq 9.4 \times 10^{24} ~{\rm m}$
which is the typical distance scale of the galaxy lensing.
\begin{figure}[htbp]
\begin{center}
 \includegraphics[scale=1.0,clip]{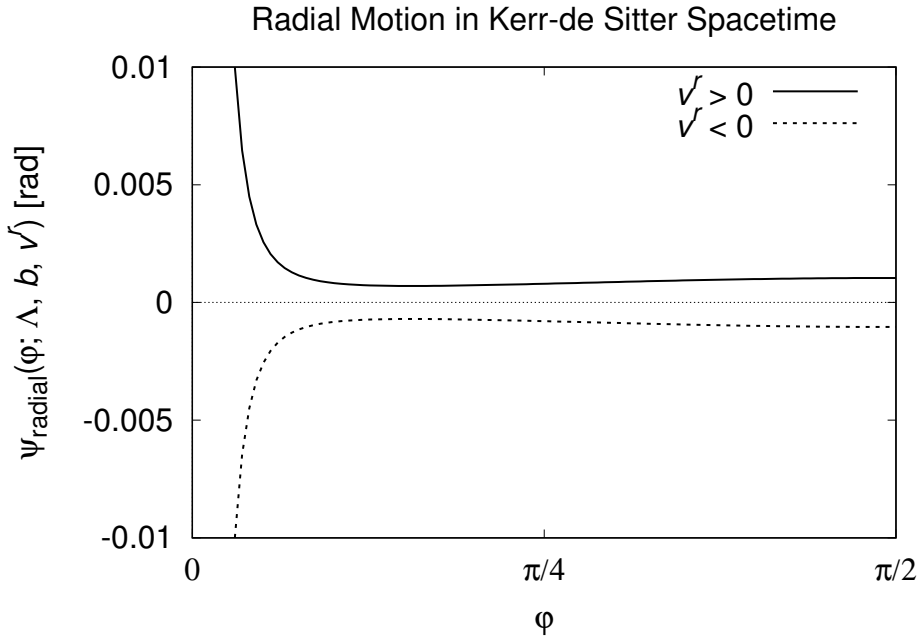}
 \caption{$\phi$ dependence of Eq. (\ref{eq:psi_radial_part1}).
 \label{fig:arakida-fig4}}
\end{center}
\end{figure}
\begin{figure}[htbp]
\begin{center}
 \includegraphics[scale=1.0,clip]{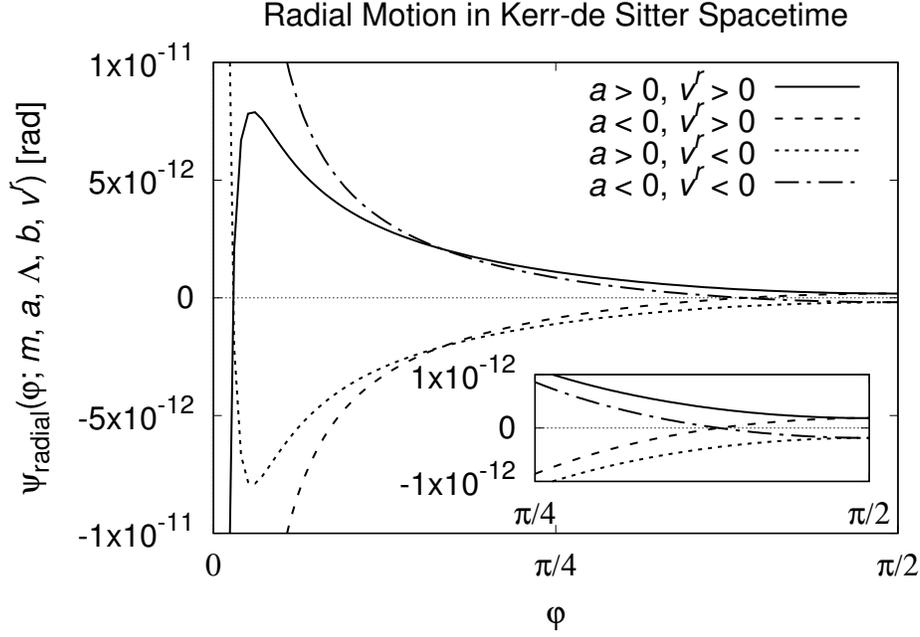}
 \caption{$\phi$ dependence of Eq. (\ref{eq:psi_radial_part2}).
 \label{fig:arakida-fig5}}
\end{center}
\end{figure}

FIGs. \ref{fig:arakida-fig4} and \ref{fig:arakida-fig5} illustrate
the $\phi$ dependence of Eqs. (\ref{eq:psi_radial_part1}) and
(\ref{eq:psi_radial_part2}), respectively.
Eq. (\ref{eq:psi_radial_part1}) diverges to positive or negative
infinity when $\phi$ approaches 0 depending on the sign of the
velocity $v^r$. On the other hand, when $\phi \rightarrow \pi/2$,
Eq. (\ref{eq:psi_radial_part1}) approaches $\Lambda b^2 v^r/3$.

For the positive velocity $v^r > 0$, Eq. (\ref{eq:psi_radial_part2})
becomes negative infinity when $\phi$ approaches 0, 
while for the negative velocity $v^r < 0$, Eq. (\ref{eq:psi_radial_part2})
diverges to positive infinity. These properties are independent of
the sign of the spin parameter $a$. Next, when $\phi \rightarrow \pi/2$,
Eq. (\ref{eq:psi_radial_part2}) converges to $\Lambda b m v^r$
which depends on the radial velocity $v^r$ but is independent of
the spin parameter $a$. The magnitude of the velocity-dependent part,
Eq. (\ref{eq:psi_radial_part2}), is at most ${\cal O}(10^{-12})$
which is two orders of magnitude smaller then the second order
contribution ${\cal O}(m^2)$ in Eq. (\ref{eq:def_angle2}).

%%%%%%%%%%%%%%%%%%%%%%%%%%%%%%%%%%%%%%%%%%%%%%%%%%%%%%%%%%%%%%%%%
\subsection{Measurable Angle by Observer in Transverse Motion}
As the third case, let us investigate the observer in transverse
motion which is the motion in a direction perpendicular
to the radial direction in the orbital plane.
The component of the 4-velocity of the observer $u^{\mu}$ is
\begin{eqnarray}
 u^{\mu} = (u^t, 0, 0, u^{\phi}),
\end{eqnarray}
and the condition $g_{\mu\nu}u^{\mu}u^{\nu} = -1$ gives
\begin{eqnarray}
 u^t = \frac{C(r)u^{\phi} +
  \sqrt{[C(r)u^{\phi}]^2 + A(r)[D(r)(u^{\phi})^2 + 1]}}{A(r)},
  \label{eq:ut_circ}
\end{eqnarray}
in which we chose the sign of $u^t$ to be positive.
$g_{\mu\nu}u^{\mu}k^{\nu}$ and $g_{\mu\nu}u^{\mu}w^{\nu}$ are
computed as
\begin{eqnarray}
 g_{\mu\nu}u^{\mu}k^{\nu}
 &=&
  \frac{A(r)D(r) + C^2(r)}{A(r)}
  \left\{
   u^{\phi}
   -
   \frac{\sqrt{[C(r)u^{\phi}]^2 + A(r)[D(r)((u^{\phi})^2 + 1)]}}
   {bA(r) - C(r)}
  \right\}k^{\phi},
   \label{eq:uk_circ1}\\
 g_{\mu\nu}u^{\mu}w^{\nu}
 &=& -
  \sqrt{
  \frac{B(r)}{A(r)}
  \left\{
   [C(r)u^{\phi}]^2 + A(r)[D(r)(u^{\phi})^2 + 1]
       \right\}}w^r.
  \label{eq:uw_circ1}
\end{eqnarray}
In the same way as was done for Eqs. (\ref{eq:uk_rad2}) and
(\ref{eq:uw_rad2}), we rewrite Eqs. (\ref{eq:uk_circ1}) and
(\ref{eq:uw_circ1}) in terms of the coordinate angular velocity
$v^{\phi}$ which is determined by
\begin{eqnarray}
 v^{\phi} = \frac{d\phi}{dt} = \frac{d\phi/d\tau}{dt/d\tau}
  = \frac{u^{\phi}}{u^t},
  \label{eq:vp1}
\end{eqnarray}
and using Eq. (\ref{eq:ut_circ}), $u^{\phi}$ is obtained by
means of $v^{\phi}$ as,
\begin{eqnarray}
 u^{\phi} = \frac{v^{\phi}}
  {\sqrt{A(r) - 2C(r)v^{\phi} - D(r)(v^{\phi})^2}}.
  \label{eq:vp2}
\end{eqnarray}
Inserting Eq. (\ref{eq:vp2}) into
Eqs. (\ref{eq:uk_circ1}) and (\ref{eq:uw_circ1}), 
$g_{\mu\nu}u^{\mu}k^{\nu}$ and $g_{\mu\nu}u^{\mu}w^{\nu}$
are rewritten as
\begin{eqnarray}
 g_{\mu\nu}u^{\mu}k^{\nu}
  &=&
  \frac{[A(r)D(r) + C^2(r)](-1 + bv^{\phi})}
  {[bA(r) - C(r)]\sqrt{A(r) - 2C(r)v^{\phi} - D(r)(v^{\phi})^2}}
  k^{\phi},
  \label{eq:uk_circ2}\\
 g_{\mu\nu}u^{\mu}w^{\nu}
  &=&
  - \sqrt{\frac{B(r)}{A(r)}}
  \frac{A(r) - C(r)v^{\phi}}
  {\sqrt{A(r) - 2C(r)v^{\phi} - D(r)(v^{\phi})^2}}w^r.
  \label{eq:uw_circ2}
\end{eqnarray}
Because $v^{\phi} = d\phi/dt$ is the coordinate angular velocity,
we regard $bv^{\phi}$ as the coordinate transverse velocity
which allows us to employ the slow motion approximation
$bv^{\phi} \ll 1$. As was done when deriving Eqs. (\ref{eq:psi_rad1})
and (\ref{eq:psi_rad1}), we substitute Eqs. (\ref{eq:Ar}), (\ref{eq:Br}),
(\ref{eq:Cr}), (\ref{eq:Dr}), (\ref{eq:kw}), (\ref{eq:uk_circ2}) and
(\ref{eq:uw_circ2}) into Eq. (\ref{eq:aberration1}),
and expand up to the order
${\cal O}(\varepsilon^2, \varepsilon^2 bv^{\phi})$,
obtaining $\psi_{\rm transverse}$ for $0 \le \phi \le \pi/2$:
\begin{eqnarray}
 \psi_{\rm transverse}
  &=&
  \phi + bv^{\phi}\tan\frac{\phi}{2}
  + \frac{2m}{b}\cos\phi\left(1 + \frac{bv^{\phi}}{1 + \cos\phi}\right)
  \nonumber\\
 & &+
  \frac{1}{8b^2}
  \left\{
   m^2[15(\pi - 2\phi) - \sin 2\phi] - 16am\cos\phi
   \right\}\nonumber\\
 & &+\frac{bv^{\phi}}{8b^2(1 + \cos\phi)}
  \Biggl\{
  m^2\left[15(\pi - 2\phi)
      - 16\sin \phi + 7\sin 2\phi + 16\tan \frac{\phi}{2}
     \right]\Biggr.
  \nonumber\\
 & &
  \Biggl.
  + 8ma(1 - 2\cos\phi - \cos 2\phi)
  \Biggr\}
  \nonumber\\
 & &- \frac{\Lambda b^2}{6}\cot\phi
  - \frac{\Lambda b^3v^{\phi}\cot\phi }{6(1 + \cos\phi)}
  + \frac{\Lambda b}{3}
  \cos\phi\left[m(1 + \csc^2\phi) + 2a\csc\phi\right] 
  \nonumber\\
 & &+ \frac{\Lambda b^3v^{\phi}
  \left[m(1  - 4\cos \phi + \cos 2\phi) -
   2a (\sin\phi + \sin 2\phi)\right]}
  {6b(\cos\phi - 1)(1 + \cos\phi)^2}
  \nonumber\\
 & &- \frac{\Lambda^2 b^4}{288}\csc^4\phi\sin 4\phi
  - \frac{\Lambda^2 b^5v^{\phi}}{2304}
  \cos\phi(1 - 2\cos\phi + 3\cos 2\phi)
  \csc^3\frac{\phi}{2}\sec^5\frac{\phi}{2}
  \nonumber\\
 & &+ {\cal O}(\varepsilon^3, (bv^{\phi})^2),
  \label{eq:psi_circ1}
\end{eqnarray}
and for $\pi/2 \le \phi \le \pi$:
\begin{eqnarray}
 \psi_{\rm transverse}
  &=&
  \pi - \phi + bv^{\phi}\cot\frac{\phi}{2}
  - \frac{2m}{b}\cos\phi\left(1 + \frac{bv^{\phi}}{1 - \cos\phi}\right)
  \nonumber\\
 & &- \frac{1}{8b^2}\{m^2[15(\pi - 2\phi) - \sin 2\phi] - 16am\cos\phi\}
  \nonumber\\
 & &+ \frac{bv^{\phi}}{8b^2(1 - \cos\phi)}
  \Biggl\{
   m^2\left[
       -15(\pi - 2\phi)
       - 16\sin\phi - 7\sin 2\phi + 16\cot\frac{\phi}{2}
      \right]\Biggr.
   \nonumber\\
 & & \Biggl.+ 8am(1 + 2\cos\phi - \cos 2\phi)
  \Biggr\}
  \nonumber\\
 & &+\frac{\Lambda b^2}{6}\cot\phi
  + \frac{\Lambda b^3 v^{\phi}\cot\phi}{6(1 - \cos\phi)}
  -  \frac{\Lambda b}{3}
  \cos\phi\left[m(1 + \csc^2\phi) + 2a\csc\phi\right] 
  \nonumber\\
 & &- \frac{\Lambda b^3 v^{\phi}
  \left[
   m(1 + 4\cos\phi + \cos 2\phi) - 2a(\sin\phi - \sin 2\phi)
	  \right]}
  {6b(\cos\phi - 1)^2(1 + \cos\phi)}
  \nonumber\\
 & &+ \frac{\Lambda^2 b^4}{288}\csc^4\phi\sin 4\phi
  + \frac{\Lambda^2 b^5 v^{\phi}}{2304}
  (1 + 2\cos\phi + 3\cos 2\phi)\csc^5\frac{\phi}{2}\sec^3\frac{\phi}{2}
  \nonumber\\
  & &+ {\cal O}(\varepsilon^3, (bv^{\phi})^2).
  \label{eq:psi_circ2}
\end{eqnarray}
As is the case of Eqs. (\ref{eq:psi_rad1}) and (\ref{eq:psi_rad2}),
the first four lines in Eqs. (\ref{eq:psi_circ1}) and
(\ref{eq:psi_circ2}) are equivalent to the measurable angle in Kerr
spacetime obtained in \cite{arakida2018b}, and the remaining terms,
lines 5 to 7, are the additional terms due to the
cosmological constant $\Lambda$.

Even here, we pay attention to the contribution of the cosmological
constant $\Lambda$ and take the parts of the cosmological constant
$\Lambda$ and the transverse velocity $bv^{\phi}$:
\begin{eqnarray}
 \psi_{\rm transverse}(\phi; \Lambda, b, bv^{\phi})
  &=&
  - \frac{\Lambda b^3v^{\phi}\cot\phi }{6(1 + \cos\phi)}
  \nonumber\\
  & &- \frac{\Lambda^2 b^5v^{\phi}}{2304}
  \cos\phi(1 - 2\cos\phi + 3\cos 2\phi)
  \csc^3\frac{\phi}{2}\sec^5\frac{\phi}{2},
  \label{eq:psi_trans_part1}\\
  \psi_{\rm transverse}(\phi; m, a, \Lambda, b, bv^{\phi})
  &=&
  \frac{\Lambda b^3v^{\phi}
  \left[m(1  - 4\cos \phi + \cos 2\phi) -
   2a (\sin\phi + \sin 2\phi)\right]}
  {6b(\cos\phi - 1)(1 + \cos\phi)^2},
  \label{eq:psi_trans_part2}
\end{eqnarray}
and we assume that the transverse velocity is comparable with
the recessional velocity $v_H$
\begin{eqnarray}
 bv^{\phi} \approx v_H,
\end{eqnarray}
see also TABLE \ref{tab:arakida-table1}.

\begin{figure}[htbp]
\begin{center}
 \includegraphics[scale=1.0,clip]{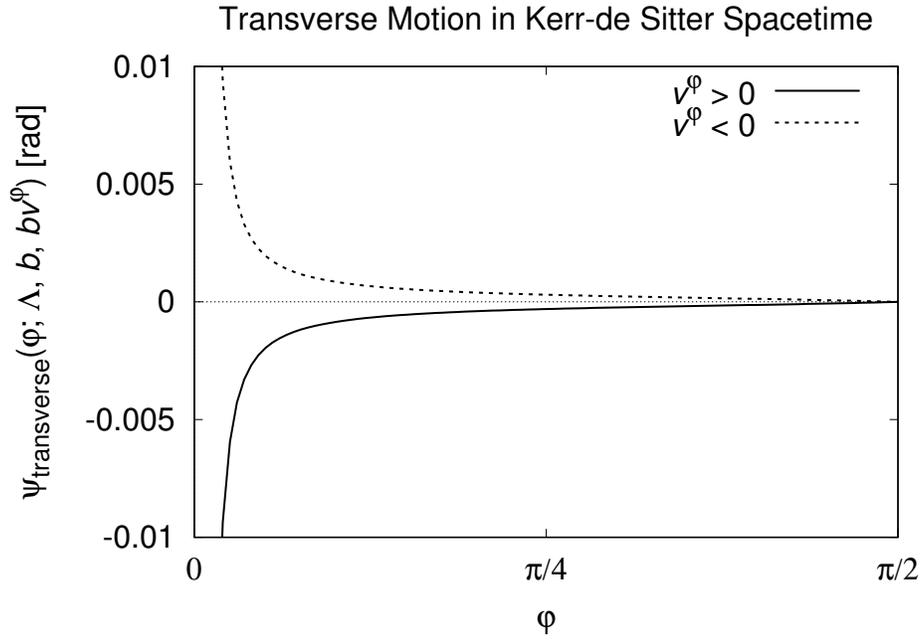}
 \caption{$\phi$ dependence of Eq. (\ref{eq:psi_trans_part1}).
 \label{fig:arakida-fig6}}
\end{center}
\end{figure}

\begin{figure}[htbp]
\begin{center}
 \includegraphics[scale=1.0,clip]{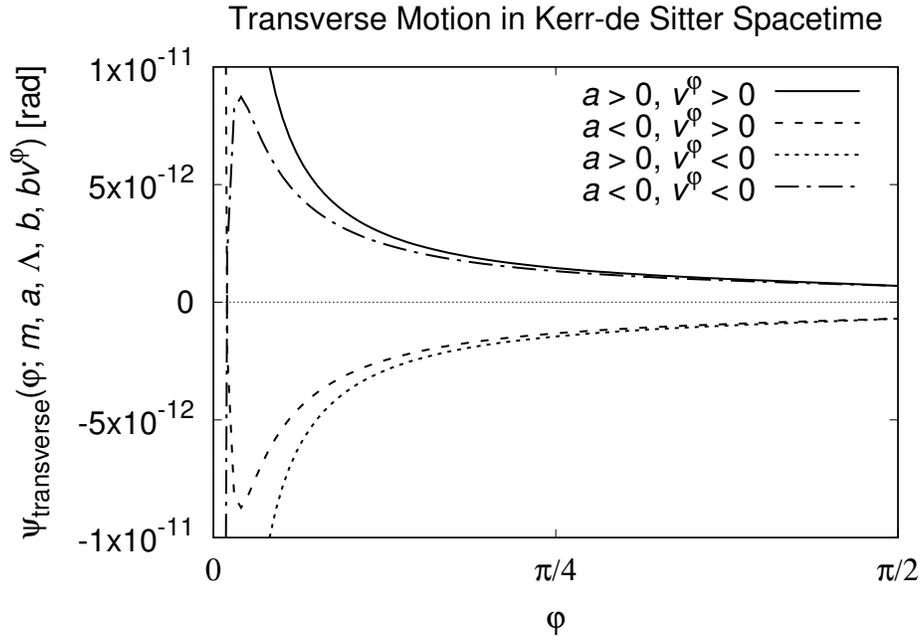}
 \caption{$\phi$ dependence of Eq. (\ref{eq:psi_trans_part2}).
 \label{fig:arakida-fig7}}
\end{center}
\end{figure}

FIGs. \ref{fig:arakida-fig6} and \ref{fig:arakida-fig7} show the
$\phi$ dependences of Eqs. (\ref{eq:psi_trans_part1}) and
(\ref{eq:psi_trans_part2}), respectively.
Eq. (\ref{eq:psi_trans_part1}) consists of the order
${\cal O}(\Lambda b^3 v^{\phi}, \Lambda^2 b^5 v^{\phi})$ terms,
however unlike Eq. (\ref{eq:psi_radial_part1}),
for the positive transverse velocity $bv^{\phi} > 0$,
Eq. (\ref{eq:psi_trans_part1}) diverges to negative infinity
when $\phi$ approaches 0 and vice versa. When
$\phi \rightarrow \pi/2$, Eq. (\ref{eq:psi_trans_part1})
converges to $0$ regardless of the sign of the transverse velocity
$bv^{\phi}$.

From FIG. \ref{fig:arakida-fig7}, we find that regardless of
the sign of the spin parameter $a$, 
Eq. (\ref{eq:psi_trans_part2}) becomes positive infinity for
the positive transverse velocity $bv^{\phi} > 0$ and negative
infinity for the negative transverse velocity $bv^{\phi} < 0$
when $\phi$ approaches 0. When $\phi \rightarrow \pi/2$,
Eq. (\ref{eq:psi_trans_part2}) converges to $\Lambda b^2 a v^{\phi}/3$
which depends on both the spin parameter $a$ and the transverse velocity
$bv^{\phi}$ unlike Eq. (\ref{eq:psi_radial_part2}).
As in Eq. (\ref{eq:psi_radial_part2}), the magnitude of
Eq. (\ref{eq:psi_trans_part2}) is at most ${\cal O}(10^{-12})$ and
it is two orders of magnitude of smaller than the second order
contribution ${\cal O}(m^2)$ in Eq. (\ref{eq:def_angle2}).

%%%%%%%%%%%%%%%%%%%%%%%%%%%%%%%%%%%%%%%%%%%%%%%%%%%%%%%%%%%%%%%%%%%
\subsection{Comparison of Static, Radial and Transverse Cases
  \label{sec:comaprison}}

Here, we summarize the asymptotic behavior, $\phi \rightarrow 0$ and
$\phi \rightarrow \pi/2$, of Eqs.
(\ref{eq:psi_sta_part1}), (\ref{eq:psi_sta_part2})
(\ref{eq:psi_radial_part1}), (\ref{eq:psi_radial_part2}),
(\ref{eq:psi_trans_part1}), and (\ref{eq:psi_trans_part2})
and their sign within the range $0 < \phi < \pi/2$
in TABLE \ref{tab:arakida-table2}.
\begin{table}[htbp]
 \caption{\label{tab:arakida-table2}
 Asymptotic behavior of Eqs.
 (\ref{eq:psi_sta_part1}), (\ref{eq:psi_sta_part2}),
 (\ref{eq:psi_radial_part1}), (\ref{eq:psi_radial_part2}),
 (\ref{eq:psi_trans_part1}), and (\ref{eq:psi_trans_part2}).}
\begin{ruledtabular}
\begin{tabular}{ccccc}
 \textrm{Motion of Observer} &
 \textrm{Eq. Number} &
 $\phi \rightarrow 0$ &
$0 < \phi < \pi/2$&
$\phi \rightarrow \pi/2$\\
 \colrule
 %First line
 \multirow{3}{*}{Static} & Eq. (\ref{eq:psi_sta_part1}) & $- \infty$ &
	     Negative & $0$  \\
 \cline{2-5}
 %Second line
 & \multirow{2}{*}{Eq. (\ref{eq:psi_sta_part2})} &
	 \multirow{2}{*}{$\infty$} & ${\rm Positive~ for} ~ a > 0$
	     & \multirow{2}{*}{$0$} \\
 %Third line
 &   & & Mostly Negative for $a < 0$ &    \\
 \hline
 %First line
 \multirow{6}{*}{Radial} &
     \multirow{2}{*}{Eq. (\ref{eq:psi_radial_part1})} &
	 $\infty$ for $v^r > 0$  & Positive for $v^r > 0$ &
		 \multirow{2}{*}{$\Lambda b^2 v^r/3$} \\
 %Second line
 &  & $- \infty$ for $v^r < 0$ & Negative for $v^r < 0$&  \\
 \cline{2-5}
 %Third line
 & \multirow{4}{*}{Eq. (\ref{eq:psi_radial_part2})} &
	 \multirow{2}{*}{$-\infty$ for $v^r > 0$}
	 & Mostly Positive for $a > 0$ & \multirow{4}{*}{$\Lambda b m v^r$} \\
 %Fourth line
 &  &  & Mostly Negative for $a < 0$ &                   \\
 \cline{3-4}
 %Fifth line
 &  & \multirow{2}{*}{$\infty$ for $v^r < 0$}  & Mostly Negative for $a > 0$ & \\
 %Sixth line
 &  &  & Mostly Positive for $a < 0$ &  \\
 \hline
 %First line
 \multirow{6}{*}{Transverse} &
     \multirow{2}{*}{Eq. (\ref{eq:psi_trans_part1})} &
	 $-\infty$ for $v^{\phi} > 0$ &
	  Negative for $v^{\phi} < 0$ & \multirow{2}{*}{$0$} \\
 %Second line
 & & $\infty$ for $v^{\phi} < 0$ & Positive for $v^{\phi} < 0$ &  \\
 \cline{2-5}
 %Third lin
 & \multirow{4}{*}{Eq. (\ref{eq:psi_trans_part2})}
     & \multirow{2}{*}{$\infty$ for $v^{\phi} > 0$}
	 &
	     Positive for $a > 0$ & \multirow{4}{*}{$\Lambda b^2 a v^{\phi}/3$} \\
 %Fourth line
 &                   & & Mostly Negative for $a < 0$ &                   \\
 \cline{3-4}
 %Fifth line
 & & \multirow{2}{*}{$- \infty$ for $v^{\phi} < 0$} &
	     Negative for $a > 0$ &                   \\
 %Sixth line
 & &  & Mostly Positive for $a < 0$ &                   \\
\end{tabular}
\end{ruledtabular}
\end{table}
We find that there is a difference in the asymptotic behavior of
the radial and transverse motions of the observer; for instance,
when $\phi \rightarrow 0$, Eq. (\ref{eq:psi_radial_part1}) diverges
to positive infinity whereas Eq. (\ref{eq:psi_trans_part1}) diverges
to negative infinity for the positive velocities $v^r > 0$
and $bv^{\phi} > 0$.
The same situation can be observed in the case of Eqs.
(\ref{eq:psi_radial_part2}) and (\ref{eq:psi_trans_part2}).

When $\phi \rightarrow \pi/2$, Eqs. (\ref{eq:psi_sta_part1}),
(\ref{eq:psi_sta_part2}), and (\ref{eq:psi_trans_part2})
converge to 0 whereas the results of Eq. (\ref{eq:psi_radial_part1})
depend on the radial velocity $v^r$. Despite the fact that
Eq. (\ref{eq:psi_radial_part2}) includes the spin parameter $a$ and
radial velocity $v^r$, it depends only on $v^r$ and is independent
of $a$. The results of Eq. (\ref{eq:psi_trans_part2}) depend on both
the transverse velocity $bv^{\phi}$ and the spin parameter $a$.

Within the range $0 < \phi < \pi/2$, the sign of
Eqs. (\ref{eq:psi_radial_part1}), (\ref{eq:psi_radial_part2}),
(\ref{eq:psi_trans_part1}), and (\ref{eq:psi_trans_part2})
depends on the sign of the velocity $v^r$, $bv^{\phi}$ and
the spin parameter $a$; e.g., for the positive velocity $v^r > 0$ and
$bv^{\phi} > 0$, these equations have a (mostly) positive value
for $a > 0$ but a (mostly) negative value for $a < 0$.

%%%%%%%%%%%%%%%%%%%%%%%%%%%%%%%%%%%%%%%%%%%%%%%%%%%%%%%%%%%%%%%%%%%%%
\section{Conclusions\label{sec:conclusions}}

In this paper, instead of the total deflection angle $\alpha$
we mainly focused on discussing the measurable angle of the light ray
$\psi$ at the position of the observer
in Kerr--de Sitter spacetime which includes the cosmological
constant $\Lambda$.
We investigated the contributions of the radial and transverse motions
of the observer which are related to the radial velocity $v^r$ and
the transverse velocity $bv^{\phi}$ as well as the influence
of the gravitomagnetic field or frame dragging described by the spin
parameter $a$ of the central object and the cosmological constant $\Lambda$.

The general relativistic aberration equation was employed to
incorporate the effect of the motion of the observer on the measurable
angle $\psi$. The expressions for the measurable angle $\psi$ derived
in this paper apply to the observer placed within
the curved and finite-distance region in the spacetime.

To obtain the measurable angle $\psi$, the equation of the light
trajectory was obtained in such a way that the background is
de Sitter spacetime instead of Minkowski spacetime. At the end of
section \ref{sec:static}, we showed that the choice of the
zeroth-order solution $u_0$ is important and a zeroth-order solution
$u_0$ in Kerr--de Sitter and Schwarzschild--de Sitter spacetimes
should be chosen in such a way that the background
is de Sitter spacetime, Eq. (\ref{eq:u0}). Further,
Eq. (\ref{eq:trajectory2}) or at least Eq. (\ref{eq:trajectory3})
should be used as the trajectory equation of the light ray.

We find that even when the radial and transverse velocities have
the same sign, their asymptotic behavior when $\phi$ approaches $0$ is
differs, and each diverges to the opposite infinity.

If we assume that the lens object is the typical galaxy,
the static terms ${\cal O}(\Lambda bm, \Lambda ba)$
in Eq. (\ref{eq:psi_sta_part2}) are basically
comparable with the second order deflection term ${\cal O}(m^2)$
but almost one order smaller than the Kerr deflection $-4ma/b^2$.
The velocity-dependent terms ${\cal O}(\Lambda bm v^r, \Lambda bav^r)$
in Eq. (\ref{eq:psi_radial_part2}) for radial motion and
${\cal O}(\Lambda b^2mv^{\phi}, \Lambda b^2av^{\phi})$ in
Eq. (\ref{eq:psi_trans_part2}) for transverse motion are at most
two orders of magnitude smaller than the second order deflection
${\cal O}(m^2)$. Therefore, if the second
order deflection term ${\cal O}(m^2)$ becomes detectable by
gravitational lensing, it may be possible to detect
the cosmological constant $\Lambda$ from the static terms
in Eq. (\ref{eq:psi_sta_part2}).

%%%%%%%%%%%%%%%%%%%%%%%%%%%%%%%%%%%%%%%%%%%%%%%%%%%%%%%%%%%%%%%%%%%%%
\appendix

 
\end{document}